\newcommand{\beq}{\begin{quote}}
\newcommand{\enq}{\end{quote}}
\newcommand{\be}{\begin{equation}}
\newcommand{\en}{\end{equation}}
\newcommand{\del}{\delta}

\documentstyle[12pt]{article}
\begin{document}
\title{ Reply to C. Tsallis's  ``Comment on Critique of
q-entropy for thermal statistics\\ 
by M. Nauenberg" 
} 
\date{May 13, 2003}
\author{Michael Nauenberg\\
Department of Physics\\
University of California, Santa Cruz, CA 95064 
}
\maketitle
\begin{abstract}
In this note it is shown  that in his ``Comments'',  Tsallis 
did not point out any flaws in the main criticism of my paper,
namely, that the  nonextensive q-entropy formalism fails to satisfy the 
zeroth law of thermodynamics. Details are also given of 
a rigorous  thermodynamic proof which demonstrates that, 
contrary to Tsallis's assertion, the application of a nonextensive formalism
to black-body radiation does not lead to the 
well known $T^4$ Stefan-Boltzman law.

\end{abstract}

\subsection*{}

In the abstract of his paper \cite{tsallis}, and in a section 
entitled ``About thermal contact between 
systems with different values of $q$ and the $0^{th}$ 
principle of thermodynamics'', Tsallis acknowledges that the
essential point of my critique \cite{michael}  is that the  q-entropy formalism 
does not give a prescription to obtain  the equilibrium temperature 
for such systems, as is required by this principle.
I paraphrased this failure of the  formalism  by stating 
that a conventional  thermometer satisfying Boltzmann-Gibbs (BG) statistics 
could not measure the temperature of a q-entropic system
with $q\neq 1$, and concluded that ``the laws of thermodynamics would 
therefore fail to have general validiy'' unless 
all systems have  the same value of $q$. Since 
there are  systems in thermal equilibrium  for which
$q=1$, corresponding to BG statistics, my conclusion  implies
that all systems {\it in thermal equilibrium} must
have $q=1$, and thus  correspond to BG statistics.  
But rather than point out supposed flaws in 
my criticism, Tsallis responded in  \cite {tsallis} by presenting the  results of 
a molecular dynamics simulation of coupled
rotators  with long range interactions,
which are known to exhibit a  quasistationary state 
below the transition temperature \cite {rufo}. Tsallis 
claimed that this state has ``{\it non} Boltzmannian statistics''
without, however, providing any  evidence, and
then  purported to show that 
a thermometer  constructed out of rotators
with nearest-neighbor interactions that satisfy BG statistics 
measures the temperature of this state. Tsallis's
temperature  is twice the mean
kinetic energy of the rotators, but this definition is only  
justified if the model  satisfies BG statistics.
However, Tsallis admits that this is not the case for the
rotator-model quasistationary state.
Moreover, the general thermodynamic definition of temperature 
(which is also introduced in the q- entropy formalism) is 
the derivative of the entropy with respect to the energy, but
neither the entropy nor this derivative was evaluated by Tsallis's 
in this simulation.
The details of this simulation were discussed in a separate 
paper \cite {tsallis1}, but not a shred  
of evidence was presented   that this 
quasistationary state  satisfies the  statistics
associated with q-entropy. In conclusion, Tsallis's   Comments \cite{tsallis}
cannot in any way be regarded as a rebuttal to  my 
main criticism of the q-entropy formalism in reference \cite{michael}.
   
As an example of one of the many unphysical results 
which have been obtained during the past decade by applying 
the q-entropy formalism 
to problems in thermal statistics, I discussed at some length in my critique
the case of black-body  radiation. In particular, I
pointed  out that this formalism  leads to a violation of the well-known
$T^4$ Stefan-Boltzmann law.  In his Comments \cite{tsallis} Tsallis responded
that it can be ``{\it trivially} shown that the
$T^4$  proportionality law remains the same for all
energy statistical  distribution (hence not only
the BG ones) as long as...for photons the distribution
depends on the light frequency $\nu$ and the appropriate
temperature $T$, only through [the ratio] $\nu/T$''. 
But as I explained in my critique, for a {\it nonextensive} 
theory both the  energy and the entropy density 
depend  not only on $\nu/T$, but 
also ({by definition})  on the volume $V$ of the cavity holding the 
black-body radiation. Otherwise the theory would be 
{\it extensive}. 
This volume dependence implies that 
these distributions must be functions not only
of a dimensionless ratio $h \nu/k_BT$, 
but also  of another dimensionles ratio $V/a(T)^3$, where $a(T)$ is 
a length parameter, characteristic of the system, which depends on the 
temperature $T$. By dimensional arguments or
from ordinary statistical mechanics we know that 
$a(T)\propto hc/k_B T$ is the mean  thermal wavelength of the
photons.  

To  show  the fallacy in  numerous publications  of  Tsallis
and others (for references, see \cite{michael}) who argue  that a {\it nonextensive} 
formalism lead to the $T^4$ Stefan-Boltzmann law, 
I give here the details of a thermodynamic derivation of this law 
in such a formalism. 
Suppose that $U/V=f(T)V^{\delta/3}$ and  
$S/V=g(T)V^{\delta/3}$, where $U$ is the
energy, $S$ is the black-body entropy in a cavity of volume $V$, 
$f(T)$ and $g(T)$
are unknown functions of $T$, and the exponent $\delta$ is a constant
which characterizes a nonextensive formalism.
Inserting  these expressions into the well known thermodynamic
relation
\be
dU=TdS-pdV,
\en
where $p$  is  Maxwell's  isotropic 
electromagnetic radiation pressure $p=(1/3)U/V$,
one obtains the equations
\be
\frac{df(T)}{dT}=T\frac{dg(T)}{dT}
\en
and 
\be
(4+\delta)f(T)=(3+\delta)Tg(T).
\en
These two equations can be readily solved
to give
\be
f(T)\propto T^{4+\delta}
\en
and 
\be
g(T) \propto T^{3+\delta}.
\en
Hence, contrary to Tsallis's assertions, this general  proof 
demonstrates  that to obtain the  Stefan-Boltzmann law,
both the energy and the entropy of black-body radiation must  
be extensive, i.e. $\del=0$.  Indeed,  Boltzmann
made this assumption in his original 1884 derivation  
of this law \cite{boltzmann}.
I would also like to point out that already in 1916,  
Einstein showed that Planck's formula for the
black-body energy density, which leads to
the Stefan Boltzmann law, can be derived  from his quantum
transition probabilities for the emission
and absorption of radiation \cite{einstein}. 
These transition probabilities  lead 
to the conventional Bose-Einstein statistics \cite{michael1}, 
and not to the modified versions which have been obtained 
by various  applications of the q-entropy formalism 
to black-body radiation (for references,
see \cite{michael}). Hence these applications are  also inconsistent 
with the laws of  quantum electrodynamics.

As I have  already indicated in the title of my paper, my
critique of the q-entropy formalism was confined only to  
its applications to systems in {\it thermal} equilibrium.  
Tsallis, however, devoted a major part of his  Comments 
\cite{tsallis} to  applications of this
formalism  to non-thermal problems such
as the logistic map, the standard
map, growth models, etc. These applications, however, are not
relevant to my critique, and therefore I will not 
discuss them here.

\end{document}